\documentclass[twocolumn,aps,prd,showpacs]{revtex4}

\usepackage{graphicx}
\usepackage[usenames]{color}
\usepackage{psfrag}
\usepackage[ps2pdf,colorlinks,bookmarks]{hyperref}
\usepackage{bm}

\definecolor{linkblue}{rgb}{0,0,0.8}
\definecolor{linkgreen}{rgb}{0,0.5,0}
\hypersetup{pdfpagemode=None, pdfstartview=FitH, linkcolor=linkblue, %
            citecolor=linkgreen, urlcolor=linkblue}

\newcommand{\ud}[2]{^{#1}_{\phantom{#1}#2}}

\def\beq{\begin{equation}}
\def\eeq{\end{equation}}
\def\bea{\setlength\arraycolsep{1.4pt}\begin{eqnarray}}
\def\eea{\end{eqnarray}}
\def\bit{\begin{itemize}}
\def\eit{\end{itemize}}
\def\nn{\nonumber}

\def\ie{{i.e.}}
\def\eg{{e.g.}}

\def\eq{Eq.~}
\def\eqs{Eqs.~}
\def\fig{Fig.~}

\def\lap{\nabla^2}

\def\ld{\left}
\def\rd{\right}
\def\blp{\boldsymbol{(}}
\def\brp{\boldsymbol{)}}
\def\ra{\rightarrow}
\def\tl{\tilde}

\def\ph{\phantom}
\def\fr{\frac}
\def\oo{\frac{1}}
\def\half{\frac{1}{2}}

\def\gam{\gamma}

\def\del{\delta}

\def\lam{\lambda}

\def\mn{{\mu\nu}}

\def\sig{\sigma}
\def\Sig{\Sigma}

\def\O{{\cal O}}

\def\pert{perturbation}
\def\perts{perturbations}

\def\lcdm{$\Lambda$CDM}

\begin{document}

\title{Can decaying modes save void models for acceleration?}

\author{James P. Zibin} 
\email{zibin@phas.ubc.ca}
\affiliation{Department of Physics and Astronomy, %
University of British Columbia, %
Vancouver, BC, V6T 1Z1  Canada}

\date{\today}

\begin{abstract}
   The unexpected dimness of Type Ia supernovae (SNe), 
apparently due to accelerated expansion driven by some form of dark energy 
or modified gravity, has led to attempts to explain the observations using 
only general relativity with baryonic and cold dark matter, but by 
dropping the standard assumption of homogeneity on Hubble scales.  In 
particular, the SN data can be explained if we live near the centre of 
a Hubble-scale void.  However, such void models have been shown to be 
inconsistent with various observations, assuming the void consists of 
a pure {\em growing} mode.  Here it is shown that models with 
significant decaying mode contribution today can be ruled out on the 
basis of the expected cosmic microwave background spectral distortion.  
This essentially closes one of the very few 
remaining loopholes in attempts to rule out void models, and strengthens 
the evidence for Hubble-scale homogeneity.
\end{abstract}


\maketitle

\section{Introduction}
\label{introsec}

   In recent years there has been considerable interest in using 
observations to try to confirm the assumption that the Universe is 
homogeneous and isotropic on the largest scales.  While isotropy 
about our own worldline follows almost directly from the observed 
isotropy of the cosmic microwave background (CMB), {\em radial} 
homogeneity is difficult to confirm on Hubble scales.  Galaxy surveys 
have found that the distribution of luminous matter to redshifts of at 
least $z \simeq 0.5$ is largely consistent with the standard 
cosmological constant plus cold dark matter (\lcdm) model (\eg, 
see~\cite{blakeetal10}).  However, radial inhomogeneity on the largest scales 
could be difficult to disentangle from redshift-dependent effects such as 
evolution, so it is not clear how much current surveys actually tell us 
about these scales.

   Given the isotropy of the CMB, the possibility of radial inhomogeneity 
might appear very contrived: it would imply that we are near 
the centre of a (nearly) spherically symmetric Universe, hence apparently 
violating the Copernican principle.  However, just such a situation 
has received much attention recently, in the context of late-time 
acceleration.  It was realized some years 
ago~\cite{tomita00,gtbgo99,celerier99} that if 
we are situated in an extremely large spherically symmetric underdensity, 
or void, the luminosity distance-redshift relation of Type Ia supernovae 
could be explained {\em without} the need for a cosmological 
constant or dark energy, or a modification of gravity.  (An earlier 
study of the potential pitfalls of living in a spherical underdensity 
while assuming a homogeneous model can be found in~\cite{mt95}.)  These 
models exchange the coincidence problem of \lcdm\ for a violation of the 
Copernican principle (although they are not free of temporal tuning 
problems~\cite{fmzs10}, nor do they address the ``old'' cosmological 
constant problem).  For a recent review of these void models and the 
various constraints on them, see, \eg, \cite{mn11}.

   General geometrical tests of large-scale homogeneity have been 
proposed (\eg, see~\cite{uce08,cbl08}).  In the context of void models 
for acceleration, the strongest and most robust results come from 
the requirement of fitting the {\em full} power spectrum of CMB temperature 
anisotropies.  It has been shown that {\em growing} mode void models 
which do fit the CMB predict a local Hubble rate far lower than 
observations indicate~\cite{zms08,mzs10,bnv10,mp10,bcf11}.  Claims have been 
made~\cite{cr10} that early radiation inhomogeneity can provide a 
loophole to this conclusion, although this appears unlikely due to the 
free streaming and rapid redshifting of the radiation~\cite{mzs10}.  
Modifying the primordial perturbation spectrum {\em might} provide a 
loophole~\cite{zms08,ns11}, although order-unity departures from scale 
invariance, and substantial fine tuning, would be required~\cite{mzs10}.

   Particularly promising tests of homogeneity are those that rely on 
the scattering of light from inside our past light cone.  Void models 
generically predict large CMB dipoles for off-centre comoving 
observers~\cite{mt95}, 
and hence should generate anisotropies via the kinetic Sunyaev-Zel'dovich 
(kSZ) effect~\cite{sz80} (as first suggested in~\cite{goodman95}).  In 
Refs.~\cite{gbh08b,yns10b}, this effect was used to put constraints 
on void models using galaxy cluster observations, and in~\cite{zs10} 
the {\em linear} kSZ effect due to all structure was found to 
essentially rule out a large class of void models.  In~\cite{zm11}, 
it was shown that strong linear kSZ constraints on (growing-mode) void models 
persisted under a fully relativistic treatment of the problem using the 
Lema\^itre-Tolman-Bondi (LTB) spacetime~\cite{lemaitre33,tolman34,bondi47}.  
However, it was stressed in~\cite{zm11} that there were many 
caveats to the linear kSZ calculations, both 
technical and relating to the poor state of {\em model-independent} 
knowledge of the local baryon fraction and free electron density power.

   Another method, which is much less susceptible to the uncertainties 
which plague the kSZ approach, involves observations of the CMB {\em 
spectral} distortions due to Compton scattering from inside the light 
cone~\cite{goodman95}.  While the kSZ approach relies on the knowledge 
of the free electron perturbation power, the calculation of the 
Compton $y$ distortion requires only {\em background} information.  
Although tight constraints on growing-mode void models using 
the Compton $y$ distortion have been presented~\cite{cs08}, more recent 
studies using a fully consistent LTB approach have found only very weak 
constraints~\cite{mzs10,zm11}.

   Even though the kSZ effect and spectral distortions both probe the 
inside of our past light cone, they have so far only been used to 
gauge possible inhomogeneity {\em on the surface} of the light cone, 
at relatively late times.  This is because of the usual restriction to 
{\em growing} modes of the LTB solution.  
Both growing and decaying modes are possible~\cite{silk77}, but decaying 
modes are usually discarded on the basis that they imply extreme 
inhomogeneity at early times, and hence appear difficult to reconcile 
with standard inflationary scenarios~\cite{z08}.  The assumption of vanishing 
decaying modes also enables tests based on structure~\cite{z08,zms08,mzs10}, 
since the growing-mode void can itself be treated as a linear perturbation 
from a Friedmann-Lema\^itre-Robertson-Walker (FLRW) spacetime at early times.  
The absence of decaying modes means that the large dipoles in the LTB 
models usually studied are mainly {\em local} to the scatterers---they can be 
considered as the result of ``peculiar velocities'' with respect to the 
CMB frame---and hence the ability of these techniques to probe the inside 
of the light cone has not been fully exploited.

   In standard cosmological models, any decaying modes are assumed 
to have negligible amplitudes by the end of inflation.  Decaying modes 
on our last scattering surface (LSS) have been observationally constrained 
to be a subdominant component~\cite{af05} (see also Ref.~\cite{cbk10} for 
an estimate of the effect of small-amplitude decaying modes on the LSS).  
In the context of void models for acceleration, however, a decaying mode 
may be localized around the observer (near the centre of spherical 
symmetry) and hence not extend to the LSS.  Nevertheless, such a decaying 
mode {\em would} be visible to a scatterer at the appropriate redshift, and 
so should be expected to generate substantial spectral distortions and kSZ 
anisotropies.

   It would be ideal to rule out decaying modes in void models on the basis 
of observations, rather than theoretical prejudices.  In addition, it 
has been suggested that the extra freedom from incorporating a significant 
decaying mode contribution today might ease the present severe constraints 
on void models~\cite{cfz09,cbk10}.  Therefore, 
in this work, full use of the power of the $y$ distortion to probe 
inside the light cone is made to study the observational implications 
of decaying modes and hence to constrain inhomogeneity at the earliest 
times.  The $y$ distortion is studied rather than the 
kSZ effect because of the above-mentioned ambiguities with the kSZ 
approach.  The simpler $y$ distortion calculations, combined with 
the current observational upper limits, will, nevertheless, 
be sufficient to rule out any contribution of decaying modes 
significant today, for all but extremely narrow decaying mode profiles.

   In Sec.~\ref{decmodesec}, the LTB model is introduced and various 
properties of decaying modes discussed.  Next, in Sec.~\ref{ycalcsec}, 
calculation procedures for the $y$ distortion are introduced, both at 
nonlinear order and under the linear approximation.  Results are 
presented in Sec.~\ref{resultssec}, before conclusions are made in 
Sec.~\ref{conclsec}.  A covariant derivation of the LTB null geodesic 
equations is presented in the \hyperref[appendix]{Appendix}.  Units are 
chosen such that $c = 1$.

\section{LTB decaying modes}
\label{decmodesec}

\subsection{General LTB solution}

   As mentioned in the Introduction, the general LTB spacetime consists 
of both a growing and a decaying mode~\cite{silk77}.  This section begins 
with a review of the isolation of these modes (based on Ref.~\cite{z08}), 
followed by a description of some technical properties of the decaying modes.

   For a spherically symmetric distribution of pressureless matter, 
Einstein's equations can be solved exactly, resulting in the LTB solution.  
The metric can be written
\beq
ds^2 = -dt^2 + \fr{Y'^2}{1 - K}dr^2 + Y^2d\Omega^2,
\label{metric}
\eeq
where a prime denotes the derivative with respect to comoving radial 
coordinate $r$, and $t$ is the proper time along the comoving worldlines.  
The function $K = K(r) < 1$ is arbitrary, and the areal radius $Y = Y(t,r)$ 
is given parametrically by
\beq
Y = \ld\{\begin{array}{ll}
{\displaystyle\fr{M}{K}(1 - \cosh\eta)}            &K < 0,\\[0.3cm]
{\displaystyle\fr{M}{K}(1 - \cos \eta)}            &0 < K < 1,\\[0.3cm]
{\displaystyle\ld(\fr{9M}{2}\rd)^{1/3}\ld(t - t_B\rd)^{2/3}}\quad&K = 0,\\
         \end{array}\rd.
\label{Ygen}
\eeq
\beq
t - t_B = \ld\{\begin{array}{ll}
{\displaystyle\fr{M}{(-K)^{3/2}}(\sinh\eta - \eta)}\quad &K < 0,\\[0.3cm]
{\displaystyle\fr{M}{K^{3/2}}(\eta - \sin\eta )}         &0 < K < 1.\\
                \end{array}\rd.\\
\label{tgen}
\eeq
Here $M(r)$ is a free radial function, which is set to $M(r)= r^3$ as a 
gauge condition (this implies that we cannot extend the solution past the 
``equator'' of a closed LTB model).

   Various physical quantities can be expressed in terms of these 
functions (see, e.g.,~\cite{z08}).  In particular,
\bea
4\pi G\rho &=& \fr{M'}{Y^2Y'},\label{rhodef}\\
\theta     &=& H_\parallel + 2H_\perp,\\
\Sig       &=& \fr{2}{3}\ld(H_\parallel - H_\perp\rd),\\
^{(3)}R    &=& \fr{2(KY)'}{Y^2Y'},\label{RK}
\eea
where $\rho$, $\theta$, and $\Sig$ are the matter density, expansion, 
and shear scalar, respectively, for the comoving worldlines.  For 
comoving shear tensor $\sig_\mn$ and radial spatial unit vector $r^\mu$ 
($r^\mu u_\mu = 0)$, we have $\Sig = \sig_\mn r^\mu r^\nu$.  $^{(3)}R$ is 
the Ricci curvature of the spatial comoving-orthogonal hypersurfaces.  
The radial and transverse expansion rates are given, respectively, by
\beq
H_\parallel = \fr{{\dot Y}'}{Y'}, \quad H_\perp = \fr{{\dot Y}}{Y},
\eeq
where the overdot denotes the derivative with respect to $t$.

   This exact solution contains another free radial function, 
$t_B = t_B(r)$, which is known as the ``bang time'' function 
since $t = t_B(r)$ implies $Y = 0$, leading to divergences in each of 
$\rho$, $\theta$, and $^{(3)}R$.  In the models of interest here, this 
will correspond to the cosmological singularity.  When $t_B(r)$ is 
not constant, it is often stated that the big bang occurs at different 
times at different locations.  However, we are of course free to 
define a new time coordinate $\tl t$ such that the big bang occurs 
``simultaneously'' at $\tl t = 0$.  The {\em physical} content 
of a varying $t_B$ is related to the divergence of covariant 
quantities such as the shear scalar, $\Sig$, as $t \ra t_B$.


   As described in~\cite{z08}, small-amplitude variations in spatial 
curvature, $^{(3)}R$, correspond to growing-mode perturbations about an 
Einstein-de Sitter (EdS) FLRW model, while appropriately small variations 
in bang time correspond to decaying modes.  Therefore, according to 
\eq(\ref{RK}), to isolate a pure decaying mode we must set $K(r) = 0$.  
The LTB solution, \eqs(\ref{Ygen}) and (\ref{tgen}), then becomes
\beq
Y = \ld(\fr{9M}{2}\rd)^{1/3}\ld(t - t_B\rd)^{2/3}.
\label{dmY}
\eeq
In the remainder of this work, only pure decaying modes will be 
considered.  Justification for this will be discussed in 
Sec.~\ref{conclsec}.  The main motivation of this study is to examine 
the effect of decaying modes that are significant today.  In Sec.~\ref{linT} 
we will see that this implies that we are interested in bang time 
fluctuations $\delta t_B(r) \sim t_0$, where $t_0$ is the proper age today.

\subsection{Properties of decaying modes}
\label{dmpropssec}

   Since significant decaying modes are not normally expected in standard 
inflationary scenarios, it will be worthwhile to explicate some of their 
counterintuitive properties.  Indeed, in the years {\em before} inflation 
was proposed, decaying modes were studied extensively, as there was no 
reason to ignore them {\it a priori.}  In particular, under the guise 
of ``delayed cores'' of the big bang, they were proposed as an 
explanation for quasars~\cite{novikov65}.  LTB decaying modes are 
generalizations of Schwarzschild white holes, which are time-reversed 
analogs of black holes.

   An important property of LTB models involves the presence of 
{\em shell crossings,} where comoving worldlines intersect and hence the 
pressure-free assumption becomes invalid.  In~\cite{hl85}, criteria were 
derived to ensure that no shell crossings occur; for the case of our 
$K(r) = 0$ pure decaying mode spacetime, they become
\beq
t_B' \le 0.
\label{tbpneg}
\eeq
This condition restricts us to situations in which the big bang 
singularity is ``delayed'' at the origin, \ie\ $t_B(r = 0) \ge t_B(r)$.  
Throughout this work, the condition (\ref{tbpneg}) will be assumed to 
hold.

   Another relevant feature of LTB decaying modes is the generic 
presence of cosmological {\em blue}shifts, and possibly {\em divergent} 
blueshifts.  In~\cite{hl84} it was shown that light rays emitted near 
the initial singularity will generically be blueshifted at late times 
when $t_B' \ne 0$.  This might mean, \eg, that an observer whose LSS 
intersected a decaying mode would observe the CMB temperature in the 
direction of the decaying mode to be {\em greater} than the actual 
recombination temperature!  In addition, in~\cite{eardley74} it was 
pointed out that under certain circumstances, divergent blueshifts 
can arise in such models, with the result that the white hole 
will be unstable and will convert into a black hole.  This blueshift 
behaviour can be understood broadly by noting that for $t_B' \ne 0$, the 
spacetime becomes shear dominated at early times.  Since the shear 
is defined to be trace free, this means that in some directions the 
comoving worldlines will be {\em contracting} near the singularity.  
In particular, the {\em radial} expansion rate $H_\parallel$ will be 
negative close enough to the singularity, leading to blueshifts for 
null rays that are approximately radial.  (A related discussion 
appears in Ref.~\cite{kl09}.)

   In the void model context, a decaying mode near the centre would 
not be directly observable by a central observer.  However, a 
scatterer at the appropriate redshift down the central observer's 
light cone {\em would} see the blueshifts due to the decaying mode, 
and hence would observe a strongly anisotropic CMB sky.  
This suggests that the scatterer would 
produce significant CMB spectral distortions (or kSZ anisotropies).  
Of course in a realistic cosmology, the dust source approximation 
of LTB breaks down at early times, when radiation becomes important.  
It is not clear how the blueshift behaviour will be modified in the 
radiation era.  However, since recombination occurs somewhat after 
radiation domination, the large 
blueshifts will be incurred during the dust era, when the LTB 
solution is reliable.  Thus the constraints from the spectral 
distortion will be sound.

\section{Calculating the spectral $y$ distortion}
\label{ycalcsec}

\subsection{The $y$ distortion}

   The $y$ distortion arises from the Compton scattering of anisotropic 
CMB radiation from inside our past light cone into our line of sight.  
In the single-scattering approximation, and in a spherically symmetric 
spacetime, it can be written as~\cite{stebbins07}
\beq
y = \fr{3}{16}\int_0^{z_{\rm re}}\!\!\!\!dz_s\fr{d\tau}{dz_s}\int_0^\pi\!\!\!
  d\xi\sin\xi\ld(1 + \cos^2\xi\rd)\ln^2\!\!\ld(\fr{T(z_s,\xi)}{T(z_s,\pi)}\rd).
\label{ygen}
\eeq
Here $\tau$ is the optical depth, and the outer integral only extends 
to the redshift of (the assumed abrupt) reionization, $z_{\rm re}$.  
$T(z_s,\xi)$ is the CMB temperature seen by a scatterer at redshift $z_s$ 
from the central observer.  Spherical symmetry implies that the temperature 
is a function of only one angle, $\xi$.  A radially outwards directed ray 
is chosen to have $\xi = 0$.  
Although in standard models we have $z_{\rm re} \simeq 10$, the 
reionization redshift could 
be somewhat different in void models.  Nevertheless, in the models studied 
here, most of the contribution to the redshift integral will come from 
$z_s \simeq 3$, so uncertainty in $z_{\rm re}$ should not affect the final 
results significantly.  The factor $d\tau/dz_s$ can be readily written 
in terms of the background quantities as~\cite{zm11}
\beq
\fr{d\tau}{dz_s}
   = \fr{\sig_{\rm T}f_b\ld(2 - Y_{\rm He}\rd)\rho_m(z_s)}
        {2m_p(1 + z_s)H_\parallel(z_s)}.
\eeq
Here $\sig_{\rm T}$ is the Thomson cross section, $f_b \equiv 
\rho_b/\rho_m$ is the baryonic to total matter fraction, 
$Y_{\rm He}$ is the helium mass fraction, and $m_p$ is the proton mass.

   In the case that the fluctuations over the sky in $T(z_s,\xi)$ are 
small, and dominated by the dipole $\beta(z_s)$, \eq(\ref{ygen}) 
becomes~\cite{mzs10}
\beq
y = \fr{7}{10} \int_0^{z_{\rm re}} dz_s\fr{d\tau}{dz_s}\beta(z_s)^2.
\label{ydip}
\eeq
We will see that this is generally not a good approximation for decaying 
mode void models.

   The best current constraints on the $y$ distortion come from the 
COsmic Background Explorer (COBE) satellite, which found 
$y < 1.5\times10^{-5}$ at $2\sig$ confidence~\cite{fixsenetal96}.

\subsection{Nonlinear calculation of CMB anisotropies}
\label{nonlinT}

   In order to evaluate the $y$ distortion expression, \eq(\ref{ygen}), 
we need to know the CMB temperature anisotropy seen by a scatterer at 
redshift $z_s$, $T(z_s,\xi)$.  To evaluate this temperature, we need 
to determine the redshifts along various (generally nonradial) null 
geodesics.  The general exact expression for redshift (\eg, see~\cite{zs08}), 
applied to a geodesic comoving congruence of worldlines, gives
\beq
1 + z(t,\xi) = \exp\ld[\int_t^{t_e}\ld(\oo{3}\theta
             + \sigma_\mn n^\mu n^\nu\rd)dt\rd].
\label{zgen}
\eeq
Here $z(t,\xi)$ is the total redshift incurred along the null 
geodesic between an arbitrary point, at proper time $t$, and the ray's 
endpoint, at $t = t_e$.  The integral is evaluated along the null 
ray.  The vector $n^\mu(t)$ is the normalized projection of the geodesic's 
tangent vector, $v^\mu$, orthogonal to the comoving timelike vector field 
$u^\mu$ (see the \hyperref[appendix]{Appendix}).  In other words, $n^\mu(t)$ 
is the spatial direction of propagation of the null ray seen by a comoving 
observer.  At the endpoint of the geodesic, $n^\mu(t_e)$ is directed at 
angle $\xi$ with respect to the radially outwards direction, \ie\ 
$r_\mu n^\mu(t_e) = \cos\xi$.

   In order to evaluate \eq(\ref{zgen}), we must first know how to determine 
the past-directed null geodesic with initial condition (IC) $\xi$ or 
$n^\mu(t_e)$, and then we must know where to stop evolving the geodesic, \ie\ 
we must know how to find the LSS.  The first problem is straightforward, 
in that we must solve the null geodesic equation
\beq
v^\mu_{;\nu}v^\nu = 0.
\label{vectorgeod}
\eeq
Extracting various components of \eq(\ref{vectorgeod}) (see the 
\hyperref[appendix]{Appendix}) gives
\bea
\fr{dt}{d\lam} &=& \gam,\label{dtdlam}\\
\fr{dr}{d\lam} &=& \fr{\gam n_r}{Y'},\\
\fr{d\theta}{d\lam} &=& \fr{L}{Y^2},\\
\fr{dn_r}{d\lam} &=& \half\gam\ld(n_r^2 - 1\rd)\ld(3\Sig n_r - \fr{2}{Y}\rd).
\label{dnrdlam}
\eea
Here $\theta$ is the standard spherical angular (comoving) coordinate of the 
photon (the second spherical angle can be chosen to be constant), $\lam$ is an 
affine parameter along the null ray, $\gam$ is defined by \eq(\ref{dtdlam}), 
$n_r \equiv n^\mu r_\mu$, and $L$ is a constant.  Finally, \eq(\ref{zgen}) 
now gives
\beq
1 + z(t,\xi) = \fr{\gam}{\gam_e}.
\label{zeval}
\eeq
The set of coupled ordinary differential equations (\ref{dtdlam}) to 
(\ref{dnrdlam}), together with the ICs (\ie\ the endpoint values) $t_e$, 
$r_e$, $\theta_e \equiv 0$, $n_{r,e} = \cos\xi$, and $\gam_e$, can then be 
solved numerically with standard techniques.

   Finding the LSS, \ie\ knowing where to stop the geodesic integration, 
necessarily involves some assumptions, since the details of recombination 
in the region of the decaying mode are unclear.  In this work, it will 
be assumed that recombination occurs at the same energy density and local 
temperature within the decaying mode as asymptotically outside.  
Schematically, the approach is to first use \eqs(\ref{dtdlam}) to 
(\ref{zeval}) to evolve a null geodesic from the centre of symmetry 
today into the past to a redshift of $z_{\rm LS} = 1091$, where 
the energy density $\rho_{\rm LS}$ is evaluated.  Then, for each scatterer 
redshift $z_s$, several null rays are propagated, again into the past, 
from the scatterer in several spatial directions as characterized by 
the initial angle $\xi$.  These null rays are evolved until the local 
energy density reaches $\rho_{\rm LS}$, at time $t_{\rm LS}(\xi)$, which 
is taken to define the LSS in that direction.  The total redshift, 
$z\blp t_{\rm LS}(\xi),\xi\brp$, incurred from the central point today, to 
the scatterer, to the LSS, is then evaluated using \eq(\ref{zeval}).  Finally, 
the temperature observed at the scatterer in direction $\xi$ can be written
\beq
T(z_s,\xi)
   = T_0\fr{(1 + z_{\rm LS})(1 + z_s)}{1 + z\blp t_{\rm LS}(\xi),\xi\brp},
\label{Tnonlin}
\eeq
where $T_0$ is the (unscattered) CMB temperature today.  This expression 
can then be used to evaluate the integral for the $y$ distortion, 
\eq(\ref{ygen}).  Note that this technique generalizes that of 
Ref.~\cite{mzs10} to the case of nonradial geodesics, and to a LSS 
determined by fixed density, rather than fixed proper time (which is a 
reasonable approximation in the case of vanishing decaying modes).  The 
dipole $\beta(z_s)$ in \eq(\ref{ydip}) is calculated in the LTB 
model by propagating past-directed null rays radially inwards and 
outwards from the scatterer to $\rho_{\rm LS}$, and comparing the 
resulting redshifts as described in~\cite{mzs10}.  

   Importantly, the uncertainties about recombination in the decaying 
mode region will likely affect the calculations of $T(z_s,\xi)$ and hence 
of $y$.  Since the temperature at recombination is determined by 
{\em local} atomic physics, the assumption that the recombination 
temperature is constant is likely reasonable.  However, it is 
certainly possible that there may be order-unity uncertainties in 
the density at recombination, due to the presence of shear close to 
the singularity.  (Likewise, as mentioned earlier, the radiation 
component absent from the LTB solution is expected to affect the 
$T(z_s,\xi)$ calculation at the level of tens of percent.)  
Nevertheless, we will see that the results presented 
here are powerful enough that order-unity variations in the density will 
not affect the final conclusions.

\subsection{Linear calculation of CMB anisotropies}
\label{linT}

   As a check of the nonlinear calculation described in 
Sec.~\ref{nonlinT}, and especially given that such calculations 
for decaying modes have apparently not been made previously, 
it will be useful to perform an alternative calculation of $T(z_s,\xi)$ 
in the case where a linear description is valid.  A decaying 
mode associated with bang time fluctuation $\del t_B(r)$ will be 
accurately described by linear theory at time $t$ when 
$\del t_B(r)/t \ll 1$~\cite{z08}.  (This explains the condition 
$\delta t_B(r) \sim t_0$ for decaying modes to have significant amplitude 
today.)  Therefore, if we satisfy 
$\del t_B(r)/t_{\rm LS} \ll 1$ at the time of recombination, then 
the spacetime will be accurately described by a linear perturbation 
from FLRW for all times after recombination (since growing modes 
are ignored).

   In particular, we can describe a small-amplitude decaying mode 
in terms of linear metric perturbations in some gauge.  A convenient 
choice is Newtonian (or zero shear) gauge, for which the vanishing of 
anisotropic stress implies that there is only one unique metric \pert, 
$\psi$, which describes both the lapse and isotropic spatial metric 
(curvature) \perts\ (\eg, see~\cite{mfb92}).  Then the temperature 
anisotropy viewed in direction $n^\mu$ is described by the Sachs-Wolfe (SW) 
effect~\cite{sw67}.  In the approximation of abrupt recombination 
(which is a good approximation on large scales), this can be 
written~\cite{zs08}
\bea
\fr{\del T(n^\mu)}{T_e}
   &=& 2\int_{t_{\rm LS}}^{t_e}\dot\psi dt
    + \oo{3}\psi - \fr{2}{9}\ld(\fr{3}{H}\dot\psi
       - \fr{\lap}{a^2H^2}\psi\rd)\nn\\
   &-& \fr{2}{3}\oo{H^2}\ld(\dot\psi + H\psi\rd)_{;\mu}n^\mu.
\label{zsgsw}
\eea
Here $a$ and $H$ are the FLRW background scale factor and Hubble rate, 
and the quantities outside the integral are to be evaluated at the 
point on the LSS in direction $n^\mu$.  $T_e$ is the background temperature 
at the null ray endpoint (normally taken to be the scatterer at $z_s$).  
(Note that \eq(\ref{zsgsw}) corrects a typographical sign error 
in~\cite{zs08}.)

   Equation~(\ref{zsgsw}) is general in that it applies to growing or 
decaying modes.  However, we can simplify it for the case of decaying 
modes.  In a dust background, the metric perturbation $\psi$ satisfies 
the equation (\eg, see~\cite{mfb92})
\beq
\ddot\psi + 4H\dot\psi = 0.\
\eeq
The decaying mode solution is
\beq
\psi(t,r) = \psi(t_e,r)\ld(\fr{t_e}{t}\rd)^{5/3},
\label{psidm}
\eeq
which implies
\beq
\dot\psi = -\fr{5}{2}H\psi.
\label{psidot}
\eeq
Substituting \eq(\ref{psidot}) into \eq(\ref{zsgsw}) gives
\bea
\fr{\del T(n^\mu)}{T_e}
   &=& -5\int_{t_{\rm LS}}^{t_e}\psi\blp t,r(t)\brp H(t)dt + 2\psi\nn\\
   &+& \fr{2}{9}\fr{\lap}{a^2H^2}\psi + \oo{H}\psi_{;\mu}n^\mu.
\label{gmsw}
\eea
In an EdS background, the radial component of the null trajectory is given by
\beq
r(t) = \sqrt{r_e^2 + \Delta r^2 - 2r_e\Delta r\cos\xi},
\eeq
where
\beq
\Delta r(t) \equiv \fr{2}{a_eH_e}\ld[1 - \ld(\fr{t}{t_e}\rd)^{1/3}\rd].
\eeq
Equation~(\ref{gmsw}) is the final expression for the decaying mode SW 
effect.  Note, in particular, that it includes an {\em integrated} SW 
component, since even in an EdS background, $\psi$ is time dependent.

   The final step is to relate the metric \pert\ $\psi(r)$ to the bang 
time fluctuation, $t_B(r)$.  To do this, first note that the 
relation
\beq
\fr{\del\rho}{\rho} = \fr{2}{3}\fr{\lap}{a^2H^2}\psi,
\label{rhopsi}
\eeq
where $\del\rho$ is the {\em comoving} density \pert, holds for both 
growing and decaying modes.  Then, expanding \eq(\ref{rhodef}) with 
\eq(\ref{dmY}) in terms of the small parameter $t_B(r)/t$, we can write
\beq
\fr{\del\rho}{\rho} = 2\ld(\fr{t_B}{t} + \fr{t_B'}{t}\fr{M}{M'}\rd)
   + \O\ld(\fr{t_B}{t}\rd)^2.
\label{rhotb}
\eeq
Note that this result generalizes the corresponding result in~\cite{z08} 
to the case of significant bang time gradients, $t_B'$.  Now, given a 
bang time function, all the pieces are in place to perform the linearized 
calculation of the CMB anisotropies, and hence of the $y$ distortion.

\section{Results}
\label{resultssec}

\subsection{Bang time profile}

   In order to calculate the $y$ distortion, we must first choose a bang 
time function.  A convenient choice is a Gaussian profile, specified by
\beq
t_B(r) = t_{B,m}e^{-r^2/L^2},
\eeq
with amplitude $t_{B,m} > 0$ and characteristic width $L$.  This 
specification is all we need to calculate the $y$ distortion according to 
the nonlinear prescription of Sec.~\ref{nonlinT}.  However, for the 
linear calculation of Sec.~\ref{linT}, we first need to determine the 
corresponding metric \pert, $\psi$.  It is straightforward to show that 
\eqs(\ref{rhopsi}) and (\ref{rhotb}) are satisfied when $\psi$ takes the form
\beq
\psi(t,r) = -\ld(a_0LH_0\rd)^2 \fr{t_{B,m}}{2t_0}\ld(\fr{t_0}{t}\rd)^{5/3}
              e^{-r^2/L^2}.
\label{psiprofile}
\eeq
Here the right-hand side has been written in terms of the quantities today, 
$a_0$, $H_0$, and $t_0$, since they are constant.  The fact that $\psi(t,r)$ 
is also a spatial Gaussian explains why the Gaussian choice for $t_B(r)$ is 
convenient.

   For the calculations presented in this section, units were chosen such 
that $100\, {\rm km} \, {\rm s^{-1}} \, {\rm Mpc^{-1}} = 1$ and 
$4\pi G = 1$.  This means that realistic values of the Hubble parameter 
today are $H_0 \simeq 0.7$, and that values of the matter density and 
proper time today are both of order unity.  It is difficult in general 
inhomogeneous cosmologies to unambiguously describe distances.  Therefore, 
the bang time profile width $L$ will be translated into a corresponding 
redshift down the central observer's light cone, $z_L \equiv z(r = L)$.  
As already mentioned, we will be primarily interested in the regime 
$t_{B,m} \sim t_0$, in which the decaying modes will have a significant 
effect at late times.  All of the calculations in this study used a value 
$H_0 = 0.7$ and integrated to a reionization redshift of $z_{\rm re} = 11$.

\subsection{Convergence to linear theory}

\begin{figure}
\includegraphics[width=\columnwidth]{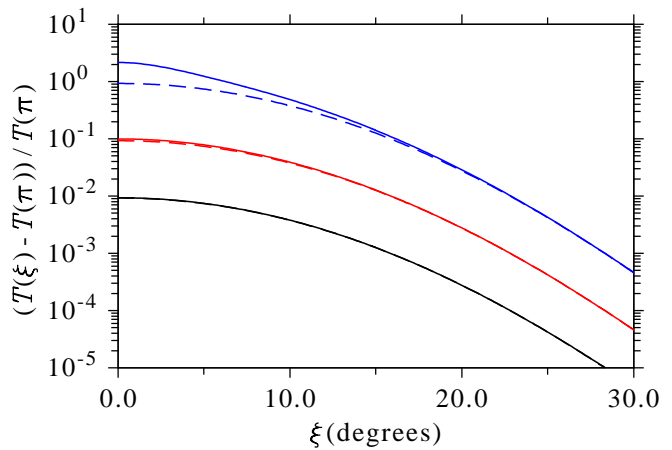}
\caption{CMB temperature anisotropy visible to a scatterer at $(t,r) = 
(1.0, 1.15)$.  The angle $\xi$ is measured from the radial direction, 
\ie\ $\xi = 0$ corresponds to the centre of the anisotropy.  Each 
solid curve represents a nonlinear calculation using 
the method of Sec.~\ref{nonlinT}, while each dashed curve used the 
linear method of Sec.~\ref{linT}.  The bang time amplitude took the 
values $t_{B,m} = 1.0\times10^{-3}$, $1.0\times10^{-4}$, and 
$1.0\times10^{-5}$, for the top (blue) curves, middle (red) curves, and 
bottom (black) curves, respectively.  The nonlinear and linear 
calculations converge at the smallest bang time amplitudes, as expected 
(the bottom two curves are almost indistinguishable).
\label{linnonlin}}
\end{figure}

   Before calculating the $y$ distortion, it will be important to check 
that the nonlinear calculation of temperature anisotropies agrees with 
the linear calculation, in the appropriate limit.  In \fig\ref{linnonlin}, 
the anisotropy $T(t_s,r_s,\xi)$ is displayed for a scatterer at position 
$(t_s,r_s) = (1.0,1.15)$, calculated using both the nonlinear and linear 
approaches.  The scatterer's position is such that the scatterer's past 
light cone intersects the decaying mode somewhat away from the peak of 
the Gaussian (although the results are qualitatively similar for {\em any} 
scatterer position).  The calculations used a profile width of $L = 0.25$, 
which corresponds to $z_L \simeq 0.35$.  As the decaying mode amplitude 
$t_{B,m}$ decreases, it is apparent that the two calculations converge, 
as expected.  This is particularly reassuring, considering how very 
different the anisotropy calculations are for linearized EdS and nonlinear 
LTB.  In addition, notice that the anisotropy becomes of order unity already 
for the relatively small amplitude $t_{B,m} = 1.0\times10^{-3}$.  For 
decaying modes of significant amplitude today, i.e.\ 
$t_{B,m} \sim t_0 \sim 1$, 
we therefore expect extreme anisotropies at the scatterers, due to the 
blueshifts mentioned previously.

   For the calculations in \fig\ref{linnonlin}, the time of last 
scattering corresponds to $t_{\rm LS} = 2.8\times10^{-5}$ (in the 
{\em linear} regime, in which $t_{\rm LS}$ is independent of position).  
Therefore, the three pairs of curves in that figure correspond to 
$t_{B,m}/t_{\rm LS} = 36.1$, $3.61$, and $0.361$, from top to bottom.  
We can see that the linear approximation is good even for ratios 
$t_{B,m}/t_{\rm LS}$ of order unity.  Thus the condition 
$t_{B,m}/t_{\rm LS} \ll 1$ given in Sec.~\ref{linT} for the validity 
of linear theory is too stringent.  To understand the reason for this, 
note that the anisotropies shown in \fig\ref{linnonlin} are small, even 
when $t_{B,m}/t_{\rm LS}$ is of order unity.  This occurs because 
of a near cancellation of the first two terms in the decaying mode 
SW effect, \eq(\ref{gmsw}).  Therefore, a metric \pert\ $\psi$ of order 
unity can produce small anisotropy, $\del T/T$.  This is in contrast to 
the familiar {\em growing} mode case, where the SW anisotropy will be of 
the same order as $\psi$.  Intuitively, a decaying mode {\em decays}, 
and hence has much less effect integrated over the null ray than a 
growing mode.  Note, finally, that this also means that it is important 
to include the gradient terms in \eq(\ref{gmsw}), even for the Hubble-scale 
decaying modes considered here.

\subsection{$y$ distortion}

\begin{figure}
\includegraphics[width=\columnwidth]{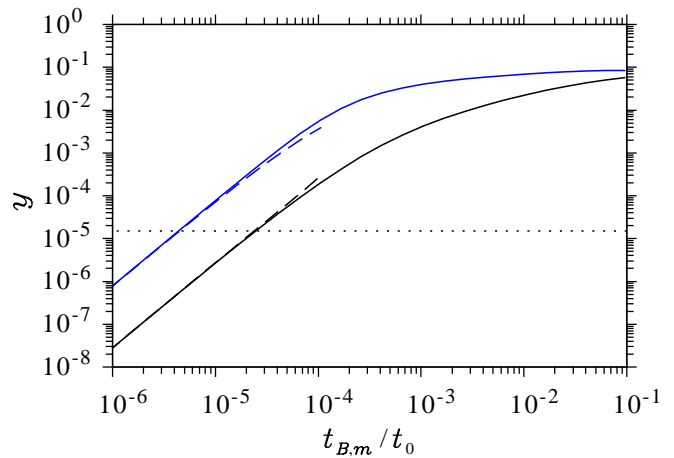}
\caption{$y$ distortion versus dimensionless bang time amplitude, 
$t_{B,m}/t_0$, at fixed profile width $L = 0.25$, which corresponds 
to $z_L \simeq 0.35$.  The black (lower) curves used the full-sky 
integration, \eq(\ref{ygen}), while the blue (upper) curves used the dipole 
approximation, \eq(\ref{ydip}).  Each solid curve used the nonlinear 
anisotropy calculation of Sec.~\ref{nonlinT}, and the dashed curves 
used the linearized SW approach of Sec.~\ref{linT}.  The nonlinear and linear 
calculations converge at the smallest bang time amplitudes, as expected.  
The dotted line indicates the COBE upper limit of $y < 1.5\times10^{-5}$ 
at $2\sig$ confidence~\cite{fixsenetal96}.
\label{ytbm}}
\end{figure}

   With all of the tools in place, we can now calculate the $y$ 
distortion.  Figure~\ref{ytbm} displays the $y$ distortion as a 
function of the dimensionless bang time profile amplitude, 
$t_{B,m}/t_0$.  These calculations also used a profile width of 
$L = 0.25$, which corresponds to $z_L \simeq 0.35$.  Curves are shown 
for the full-sky integration based 
on \eq(\ref{ygen}), together with the dipole approximation, 
\eq(\ref{ydip}).  Also, anisotropies are calculated using the nonlinear 
approach of Sec.~\ref{nonlinT}, as well as the linearized SW approach 
from Sec.~\ref{linT}.  It is evident, again, that the nonlinear calculation 
approaches the linear one for small bang time amplitudes, and that both 
exhibit the expected quadratic dependence of $y$ on $t_{B,m}/t_0$ in the 
linear regime.  It is clear 
as well that the dipole approximation substantially overestimates the 
spectral distortion, especially at the lower amplitudes.  Most striking, 
however, is the fact that the $y$ distortion exceeds the COBE limit by 
orders of magnitude, for decaying modes which are significant today 
(\ie\ for $t_{B,m}/t_0 \sim 1$), at least for decaying modes with width 
$z_L \simeq 0.35$.

\begin{figure}
\includegraphics[width=\columnwidth]{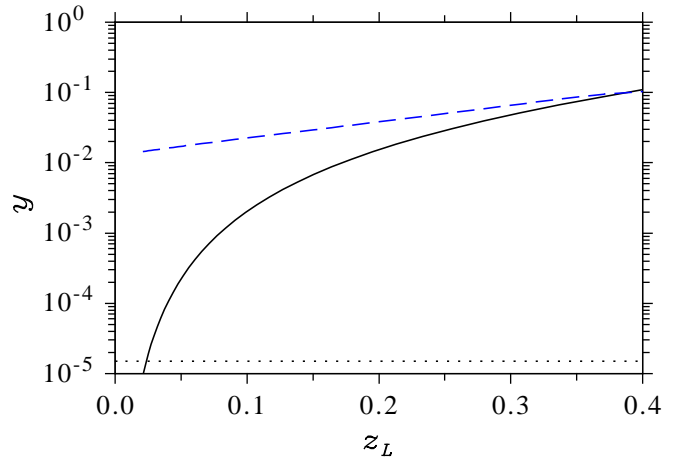}
\caption{$y$ distortion versus decaying mode width, $z_L$, at fixed bang 
time amplitude, $t_{B,m}/t_0 = 0.2$.  The black (solid) curve used the 
full-sky integration, \eq(\ref{ygen}), while the blue (dashed) curve used 
the dipole approximation, \eq(\ref{ydip}).  The dotted line indicates the 
COBE upper limit of $y < 1.5\times10^{-5}$ at $2\sig$ 
confidence~\cite{fixsenetal96}.
\label{yL}}
\end{figure}

   We should expect that the $y$ distortion will decrease as the decaying 
mode profile width decreases, due to the smaller solid angle sourcing the 
angular integral in \eq(\ref{ygen}).  Also, the anisotropies themselves 
should be affected 
[\eg\ via the $L^2$ factor in \eq(\ref{psiprofile})], although it is not 
clear exactly what $L$ dependence is expected, due to the subtle cancellations 
that occur in the SW expression, \eq(\ref{gmsw}).  In \fig\ref{yL}, the 
$y$ distortion is plotted versus decaying mode width, $z_L$, at fixed bang 
time amplitude, $t_{B,m}/t_0 = 0.2$.  These calculations used the nonlinear 
approach of Sec.~\ref{nonlinT}.  A marked decrease in $y$ 
as $z_L$ decreases is visible, with $y$ dropping below the COBE limit for 
$z_L \lesssim 0.02$.  Also notable is the fact that the dipole approximation 
severely underestimates the drop in $y$.  This is understandable, since 
the dipole approximation misses the solid angle effect mentioned above.  
Finally, note that, for the profiles able to evade the COBE constraint 
($z_L \lesssim 0.02$), the profile width must be very small 
relative to cosmological scales.

\section{Conclusions}
\label{conclsec}

   As discussed in Sec.~\ref{dmpropssec}, cosmological decaying modes suffer 
from a variety of problems.  They should have decayed to negligible amplitudes 
by the end of inflation, and no mechanism has been proposed for the generation 
of decaying modes so large as to still be significant today.  They must 
satisfy the condition $t_B' \le 0$ to avoid shell crossings, although 
there is something oddly restrictive about this criterion: clearly a generic 
profile in the absence (or even in the presence) of spherical symmetry will 
violate it somewhere.  Of course, without an explicit mechanism of 
formation, not much else can be said.  The possibility of blueshift 
instability would also need to be addressed for these models.

   In addition, if we are to save void models for acceleration by adding 
a significant decaying mode contribution today, then the ICs would need 
to be exquisitely finely tuned.  In the {\em linear} regime, the metric 
perturbation $\psi$ is constant in the growing mode, but obeys 
\eq(\ref{psidm}) for the decaying mode.  Therefore the ratio between 
growing and decaying modes increases like $t^{5/3}$.  A ratio of order 
unity today implies that the growing modes must be suppressed by a 
factor $(t_{\rm LS}/t_0)^{5/3} \sim 10^{-8}$ at last scattering.  Of course 
this estimate will be quantitatively affected by the breakdown of linear 
theory at early times, but the conclusion will still hold qualitatively:  
In a model with significant decaying modes today, the extremely 
inhomogeneous early Universe must be pure decaying mode to very high 
precision.

   One further problem with decaying mode models is related to the 
shear domination at early times.  Since the details of big bang 
nucleosynthesis (BBN) depend on the expansion history of the early 
Universe, we can expect substantial departures from standard BBN in 
these models.  Since the decaying mode profile must be centred near our 
own worldline, we would not expect the observed local light-element 
abundances to agree with predictions based on the standard \lcdm\ model.

   Nevertheless, in this study decaying modes have been given the ``benefit 
of the doubt,'' in that it has been assumed that all of these problems 
could be somehow overcome, and the observational consequences of decaying 
modes have been examined.  As mentioned in the \hyperref[introsec]
{Introduction}, while {\em growing} mode LTB void models for acceleration 
have been essentially ruled out on the basis of ${\rm CMB} + H_0$, kSZ, and 
other observations, it might be possible that a significant contribution 
today of LTB decaying mode could allow such models to survive those tests.  
The generic presence of blueshifts near the cosmological singularity 
of a decaying mode, however, suggests that these inhomogeneous bang 
time models might be susceptible to probes of the interior of our past 
light cone.  Therefore, a procedure for calculating CMB temperature 
anisotropies for a scatterer whose LSS crosses a decaying 
mode was developed and used to calculate the $y$ distortion in these 
models.  Although the results will necessarily be dependent upon the 
uncertainties regarding recombination inside a decaying mode, the 
constraints were so strong that even order-unity uncertainties in the 
calculated anisotropies will have negligible effect on the conclusions.  
Decaying modes which are significant today and wider than $z_L \simeq 0.02$ 
are ruled out by the $y$ distortion test.  Narrower profiles might 
survive this test, but would almost certainly suffer from obvious 
problems in local surveys: the region inside $z_L \simeq 0.02$ would be 
{\em much} younger than the outside, and hence would have a substantially 
higher density and expansion rate.  In addition, structure would 
presumably have substantially less power on the inside.  The local 
Universe to $z \simeq 0.02$ is well surveyed, and no such discontinuities 
are apparent.

   It is worth stressing that decaying modes with sufficiently small 
amplitudes, \ie\ $t_{B,m}/t_0 \lesssim 10^{-5}$, are not ruled out by 
the $y$ distortion test.  Although such perturbations may suffer some 
of the problems with decaying modes listed above, it is conceivable that 
they play a role in cosmology~\cite{af05}.  However, the effects of decaying 
modes with such small amplitudes will be negligible today, and hence will not 
ease the severe constraints on growing-mode void models for acceleration.

   Note that the philosophy of this study has been to examine the 
spectral distortions due to {\em pure} decaying modes.  Of course one 
could argue that a model which evades the ${\rm CMB} + H_0$ and kSZ 
tests might consist of a superposition of decaying and growing modes.  
However, given that pure growing-mode models typically produce values of 
the $y$ distortion below the COBE limit~\cite{zm11}, it appears very 
likely that the addition of growing modes of typical amplitudes will 
not change the conclusions of this study significantly.  Of course the 
combined effect of growing and decaying modes will generally not be a 
simple linear superposition, due to the nonlinearity of general 
relativity.  But the extremely large anisotropies in decaying mode models 
are due fundamentally to the extreme inhomogeneity at early times, when 
a growing-mode contribution will be negligible.

   Therefore, it appears that decaying modes do not provide a loophole 
to the conclusion that void models cannot explain acceleration.  
This result strengthens our confidence in the standard paradigm of 
a homogeneous, isotropic, accelerating cosmology, driven by some 
mysterious form of dark energy of modified gravity.

\begin{acknowledgments}
I thank Albert Stebbins and Andrzej Krasi\'nski for useful discussions.  
This research was supported by the Canadian Space Agency.
\end{acknowledgments}

\ph{blank\\}

   {\it Note added.---}After this work was essentially complete, a closely 
related paper appeared~\cite{bcf11}, which also studied the observational 
consequences of decaying modes which are significant today.  There are 
a few major differences to our approaches.  First, Ref.~\cite{bcf11} 
did not exclude the ``advanced'' big bang case, $t_B' > 0$.  This case 
was excluded in the present study since the consequent early shell 
crossings were argued to render the early-time LTB solution, and hence the 
temperature anisotropy calculations, unreliable.  Another difference is 
that Ref.~\cite{bcf11} used the kSZ effect in clusters as the basis of its 
constraints, as opposed to the $y$ distortion used here.  However, 
Ref.~\cite{bcf11} used the dipole approximation for the anisotropies seen 
by the scatterers.  Finally, the models studied in~\cite{bcf11} included 
LTB growing modes.  The strong findings of~\cite{bcf11} are in agreement 
with those of the present study, and hence reinforce the conclusion that 
models with significant decaying modes today are not viable.

\appendix*
\section{Covariant null geodesic equations}
\label{appendix}

\subsection{$1 + 1 + 2$ formalism}

   The nonradial null geodesic equations in the LTB spacetime have been 
derived many times before (e.g., see~\cite{hl84,pp90,afms93,hmm97}).  In this 
appendix, a novel derivation is presented which is based on the covariant 
$1 + 1 + 2$ formalism, and, in particular, makes no use of Christoffel symbols.

   This subsection will begin with a brief summary of the covariant 
$1 + 3$ formalism.  More details can be found in the 
reviews~\cite{tcm08,ee98}.  This formalism is based upon a fundamental 
timelike vector field, $u^\mu$ (normalized according to $u^\mu u_\mu = -1$), 
which is conveniently taken to be tangent to the dust comoving worldlines 
in the LTB spacetime.  The tensor
\beq
h\ud{\mu}{\nu} \equiv \del\ud{\mu}{\nu} + u^\mu u_\nu
\label{sliceproj}
\eeq
projects orthogonal to $u^\mu$.  We can use $h_\mn$ to define a spatial 
covariant derivative according to
\beq
D^{}_\mu T_{\nu_1\nu_2\cdots\nu_n}
   \equiv h\ud{\lam}{\mu} h\ud{\sig_1}{\nu_1}h\ud{\sig_2}{\nu_2}\cdots
          h\ud{\sig_n}{\nu_n} T_{\sig_1\sig_2\cdots\sig_n;\lam},
\label{slicecovder}
\eeq
for any tensor $T_{\nu_1\nu_2\cdots\nu_n}$ orthogonal to $u^\mu$ in all 
of its indices.

   For the case of twist-free dust, which is the case for the LTB model, 
we can decompose the covariant derivative of $u_\mu$ according to
\beq
u_{\mu;\nu} = \oo{3}\theta h_{\mn} + \sig_{\mn}.
\label{ucovder}
\eeq
The scalar $\theta$ measures the local volume rate of expansion of the 
congruence $u^\mu$.  The trace-free, symmetric tensor $\sig_{\mn}$ measures 
the local rate of shear of the congruence, and is orthogonal to 
$u^\mu$ in both of its indices.  For a dust source, the acceleration of the 
comoving worldlines vanishes, i.e.\ $u_{\mu;\rho}u^\rho = 0$.

   Under spherical symmetry, each comoving-orthogonal slice contains a 
preferred spacelike congruence with radial tangent vector $r^\mu$, where 
$r^\mu r_\mu = 1$.  In this case, the $1 + 3$ covariant approach can be 
generalized to the so-called $1 + 1 + 2$ approach, which incorporates 
both $u^\mu$ and $r^\mu$ as fundamental fields~\cite{vee96,cb03,clarkson07}.  
By analogy with the tensor $h\ud{\mu}{\nu}$ defined in \eq(\ref{sliceproj}), 
which projects into the slices, we can define a tensor $s\ud{\mu}{\nu}$ by
\beq
s\ud{\mu}{\nu} \equiv h\ud{\mu}{\nu} - r^\mu r_\nu,
\label{sheetproj}
\eeq
which projects into the two-spheres (called {\em sheets}) orthogonal to 
both $u^\mu$ and $r^\mu$.

   Under spherical symmetry, a spatial three-tensor such as the shear 
$\sig_\mn$ must be expressible in terms of $r^\mu$, $s^\mn$, and a two-scalar 
$\Sig$.  Explicitly, we can write
\beq
\sig_\mn = \ld(r_\mu r_\nu - \half s_\mn\rd)\Sig.
\label{sig112}
\eeq
This implies that the shear scalar can be written $\Sig = \sig_\mn r^\mu r^\nu$.

\subsection{Null geodesics}

   The tangent vector $v^\mu$ to a null geodesic can be decomposed into 
components parallel and orthogonal to the dust four-velocity $u^\mu$ according 
to
\beq
v^\mu = \gam(u^\mu + n^\mu),
\label{vun}
\eeq
where the spatial propagation direction $n^\mu$ satisfies $n^\mu u_\mu = 0$ 
and $n^\mu n_\mu = 1$.  By virtue of the relation $v^\mu u_\mu = -\gam$, 
the photon energy (or blackbody spectrum temperature) is proportional to 
$\gam$.  For any null geodesic in a spherically symmetric background, $n^\mu$ 
will be constrained to a two-dimensional subspace of each comoving slice.  
In other words, $n^\mu$ can be decomposed into components parallel to 
$r^\mu$ and parallel to an angular direction, denoted $\theta^\mu$, where 
$\theta^\mu\theta_\mu = 1$ and $\theta^\mu u_\mu = \theta^\mu r_\mu = 0$.  
Explicitly, 
\beq
n^\mu = n_r r^\mu + n_\theta\theta^\mu.
\eeq
Using \eq(\ref{sig112}) we can calculate the following projections 
of the shear tensor:
\beq
\sig_\mn r^\mu n^\nu = \Sig n_r, \quad
\sig_\mn n^\mu n^\nu = \half\Sig\ld(3n_r^2 - 1\rd).
\label{shearproj}
\eeq

   The null geodesic equation can be written
\beq
v\ud{\mu }{;\nu}v^\nu = \fr{dv^\mu}{d\lam} = 0,
\eeq
where $\lam$ is an affine parameter along the geodesic.  Our goal will be 
to evaluate the various components of the geodesic equation.  For the time 
component, using \eqs(\ref{ucovder}) and (\ref{shearproj}) we find
\beq
\fr{d\ld(v^\mu u_\mu\rd)}{d\lam} = -\fr{d\gam}{d\lam}
   = \oo{3}\gam^2\theta + \half\gam^2\Sig\ld(3n_r^2 - 1\rd).
\label{geodt}
\eeq
To extract the remaining components, we can first expand the geodesic 
equation using \eqs(\ref{vun}), (\ref{ucovder}), and (\ref{geodt}) to obtain
\beq
\fr{dn^\mu}{d\lam} = \oo{3}\gam\theta u^\mu - \gam\sig\ud{\mu}{\nu}n^\nu
   + \half\gam\Sig\ld(3n_r^2 - 1\rd)(u^\mu + n^\mu).
\label{dndlam}
\eeq
Then we can write the radial component as
\bea
\fr{d\ld(v^\mu r_\mu\rd)}{d\lam} &=& \fr{d\ld(\gam n^\mu r_\mu\rd)}{d\lam}\\
   &=& \fr{d\gam}{d\lam}n_r + \gam\fr{dn^\mu}{d\lam}r_\mu
       + \gam n^\mu\fr{dr_\mu}{d\lam}.
\label{dvrdlam}
\eea
The third term on the right-hand side of \eq(\ref{dvrdlam}) can be evaluated 
as follows:
\bea
n^\mu\fr{dr_\mu}{d\lam} &=& \gam n^\mu r_{\mu;\nu}\ld(u^\nu + n^\nu\rd)\\
   &=& \gam n_\theta\theta^\mu r_{\mu;\nu}n_\theta\theta^\nu\label{ndrdlam2}\\
   &=& \gam n_\theta^2\theta^\mu\theta^\nu D_\nu r_\mu\\
   &=& \half\gam n_\theta^2\theta_s.\label{ndrdlam4}
\eea
Here expression (\ref{ndrdlam2}) follows from the normalization of $r^\mu$ 
and $n^\mu$, spherical symmetry, and \eq(\ref{ucovder}).  Expression 
(\ref{ndrdlam4}) uses the normalization of $n^\mu$ and the definition
\beq
\theta_s = D_\mu r^\mu
\eeq
for the ``sheet expansion'' $\theta_s$, which measures the spatial rate of 
expansion of the spatial congruence $r^\mu$ on each comoving time slice.  
Combining \eqs(\ref{geodt}), (\ref{dndlam}), and (\ref{ndrdlam4}), we finally 
obtain the radial component of the geodesic equation:
\beq
\fr{d\ld(v^\mu r_\mu\rd)}{d\lam}
   = -\gam^2n_rH_\parallel + \half\gam^2\theta_sn_\theta^2.
\label{geodr}
\eeq

   We can obtain the final $\theta^\mu$ component of the geodesic equation 
in a completely analogous manner to the radial component.  However, we can 
do this more easily by taking the derivative with respect to $\lam$ of the 
identity
\beq
\gam^2 = \gam^2n_r^2 + \gam^2n_\theta^2
       = \ld(v^\mu r_\mu\rd)^2 + \ld(v^\mu\theta_\mu\rd)^2.
\eeq
After substituting \eqs(\ref{geodt}) and (\ref{geodr}), the result is
\beq
\fr{d\ld(v^\mu\theta_\mu\rd)}{d\lam}
   = -\gam^2n_\theta H_\perp - \half\gam^2\theta_s n_r n_\theta.
\label{geodth}
\eeq

   We now have all components of the null geodesic equation in covariant 
$1 + 1 + 2$ notation.  Our final step will be to express these equations 
in terms of coordinates along a null ray.  The most natural choice 
of coordinates is the comoving, proper time coordinates implied by the 
LTB metric, \eq(\ref{metric}).  The single nontrivial angular coordinate 
will be taken to be the standard spherical coordinate $\theta$ ($\phi$ 
will be constant).  The link between the covariant geodesic equation and 
the coordinates $x^\mu$ is through the relation $v^\mu = dx^\mu/d\lam$.  
In component form, we have
\bea
v^\mu u_\mu      &=& -\fr{dt}{d\lam},\label{vucomp}\\
v^\mu r_\mu      &=& \fr{Y'}{\sqrt{1 - K}}\fr{dr}{d\lam},\\
v^\mu \theta_\mu &=& Y\fr{d\theta}{d\lam}.\label{vthcomp}
\eea
Inserting \eqs(\ref{vucomp}) to (\ref{vthcomp}) into the geodesic equation 
components, \eqs(\ref{geodt}), (\ref{geodr}), and (\ref{geodth}), we 
finally obtain the coupled set of ordinary differential equations
\bea
\fr{dt}{d\lam} &=& \gam,\label{adtdlam}\\
\fr{dr}{d\lam} &=& \gam n_r\fr{\sqrt{1 - K}}{Y'},\\
\fr{d\theta}{d\lam} &=& \fr{\pm\gam\sqrt{1 - n_r^2}}{Y},\\
\fr{d^2t}{d\lam^2} &=& -\gam^2\ld(H_\perp + \fr{3}{2}n_r^2\Sig\rd),\\
\fr{d^2r}{d\lam^2} &=&
    -\ld(\fr{dr}{d\lam}\rd)^2\ld(\fr{Y''}{Y'} + \fr{K'}{2(1 - K)}\rd)\nn\\
   &-& 2\gam\fr{dr}{d\lam}H_\parallel
    + \ld(\fr{d\theta}{d\lam}\rd)^2\fr{Y(1 - K)}{Y'},\\
0 &=& \fr{d}{d\lam}\ld(Y^2\fr{d\theta}{d\lam}\rd),\label{angmom}\\
\fr{dn_r}{d\lam} &=& \half\gam\ld(n_r^2 - 1\rd)
                     \ld(3\Sig n_r - \fr{2\sqrt{1 - K}}{Y}\rd).\label{adnrdlam}
\eea
Here the expression $\theta_s = 2\sqrt{1 - K}/Y$ has been used.  
Equation~(\ref{angmom}) represents the conservation of angular momentum.  It 
is straightforward to verify that these equations are equivalent to those 
derived by more familiar techniques (\eg, \cite{aa06}).  The set 
(\ref{adtdlam}) to (\ref{adnrdlam}) is overdetermined, and experimentation 
revealed the subset (\ref{dtdlam}) to (\ref{dnrdlam}) (specialized to the 
decaying mode case of $K(r) = 0$) to be the most robust and numerically 
efficient.

\bibliography{bib}

\end{document}